\documentclass[12pt]{article}
\usepackage{amssymb}
\usepackage{amsmath, amsthm}
\newcommand{\be}{\begin{equation}}
\newcommand{\ee}{\end{equation}}
\begin{document}
\def\theequation{\arabic{section}.\arabic{equation}}
\begin{titlepage}
\title{Black hole entropy in scalar-tensor and $f(R)$ gravity: an 
overview}
\author{Valerio Faraoni\\ \\
{\small \it Physics Department, Bishop's University}\\
{\small \it 2600 College St., Sherbrooke, Qu\'{e}bec, Canada 
J1M~1Z7}\\
{\small \it email vfaraoni@ubishops.ca}
}
\date{} \maketitle
\thispagestyle{empty}
\vspace*{1truecm}
\begin{abstract}
A short overview of black hole entropy in alternative 
gravitational theories is presented. Motivated by the recent 
attempts to 
explain the cosmic acceleration without dark energy, we focus on 
metric and Palatini $f(R)$ gravity and on scalar-tensor theories. 
\end{abstract} 
\vskip1truecm
\begin{center}
To appear in ``Entropy in Quantum Gravity'', 
special issue  of {\em Entropy}, R. Garattini editor. 
\end{center}

\end{titlepage} 
\clearpage 
\setcounter{page}{2} 

\section{Introduction}

Alternative theories of gravity have been studied for a long 
time \cite{Willbook, Schmidtreview, Hehl76, FujiiMaeda03, 
mybook}. 
While there is no experimental evidence of deviations 
from general relativity at Solar System scales, from the 
theoretical point of view it is  well known that corrections to 
the  Einstein-Hilbert action (such as non-linear terms in the 
Ricci and Riemann tensors, or extra scalar fields coupling 
explicitly to gravity)  always arise from attempts 
to renormalize general relativity, to build a theory of quantum 
gravity or at  least some effective action (including the 
low-energy limit of string theories), or  simply from the 
quantization of scalar fields in curved 
spacetimes \cite{CCJ}, or even from purely classical arguments 
\cite{ChernikovTagirov68, SonegoFaraoni93}. From 
the observational point of view, the 1998 discovery that the 
cosmic expansion accelerates today \cite{SN1, SN2, SN3, SN4, 
SN5, SN6, SN7, SN8, SN9} makes it necessary, 
if one 
wants to remain within the bounds of general relativity, that 
96\% of the energy content of the universe be in the  
exotic form of dark energy with equation of state $P\sim 
-\rho$ (where $\rho$ and $P$ are the energy density and 
pressure of the cosmic fluid, respectively), unless one wants to 
admit a cosmological constant  of incredibly small magnitude and 
then face the cosmological 
constant problem(s)  (for a bibliography on dark energy 
see Ref.~\cite{Linderresletter}). Such explanations are 
unpalatable to 
many authors and this fact has provoked a revival of interest in 
alternative gravity theories: perhaps the explanation of the 
cosmic acceleration lies in the fact that we do not understand 
gravity at the largest scales. This possibility has led to 
extensive recent literature on metric $f(R)$ gravity 
\cite{CCT1, CCT2, CDTT}, Palatini modified  gravity 
\cite{Vollick}, and 
the  metric-affine version of these theories 
\cite{metricaffine1, metricaffine2, metricaffine3, 
metricaffine4, metricaffine5} 
(see  \cite{review, DeFeliceTsujikawa10} for reviews and 
\cite{otherreviews1, otherreviews2, otherreviews3, 
otherreviews4, otherreviews5, otherreviews6, Schmidtreview} for 
shorter  introductions). Metric and Palatini (but not 
metric-affine) modified gravities 
can be reduced to scalar-tensor theories \cite{BransDicke, ST1, 
ST2, ST3}, 
as explained below.

Black hole thermodynamics \cite{Waldbook, WaldLivRev, Waldreview} 
constitutes 
an important development of modern theoretical physics and one of 
the 
main motivations  for its study is the hope that  something will 
be learned about quantum gravity and the construction of the 
statistical mechanics underlying  this macroscopic   
thermodynamics. Black hole thermodynamics extends well beyond 
Einstein's theory of gravity; indeed, if there is hope to learn 
about quantum gravity by studying black hole 
thermodynamics, it will be necessary to understand it in 
extensions of Einstein's theory given that quantum corrections, 
renormalization, effective theories, and the low-energy limit and 
string theories all introduce extra  
gravitational scalar fields, non-minimal couplings with the 
curvature, and higher derivative corrections to general 
relativity. It is also possible that the stability of black hole 
thermodynamics with respect to perturbations of the 
Einstein-Hilbert action selects preferred classes of theories 
\cite{JacobsonKangMyers94}. In this sense, black holes can be 
regarded as ``theoretical laboratories'' for quantum gravity.

The thermodynamics of black hole (apparent and event) horizons in 
general relativity inspired the construction of a thermodynamics of 
spacetime by Jacobson using local Rindler horizons and assuming 
the 
entropy-area relation $S=\frac{A}{4G}$, where $G$ is Newton's 
constant (we use units in which the speed of light $c$  
and the reduced Planck constant $\hbar$ assume 
the value unity) \cite{Jacobson95}. Jacobson 
was able to  derive the Einstein  equations from such a 
macroscopic description 
\cite{Jacobson95}, showing that the field equations of GR are 
akin to a  macroscopic effective equation 
of state. The obvious implication of this derivation is that, if 
this picture is correct, quantizing the  Einstein equations 
would make little sense, the same way that  it would make no 
sense  to quantize the equation of state of  a (monoatomic) 
hydrogen gas in  order  to learn about the quantum states of the 
hydrogen atom.

The derivation of the field equations using the thermodynamics of 
local Rindler horizons has been performed also for metric $f(R)$ 
gravity \cite{Eling:2006aw}. In this derivation, $f(R)$ 
corrections to the Einstein-Hilbert action appear to describe 
non-equilibrium thermodynamics \cite{Eling:2006aw} (but see also 
\cite{ElizaldeSilva08, ChircoLiberati10}). Viewing the field 
equations as macroscopic equations and gravity as an emergent 
phenomenon is possible also in Lanczos-Lovelock and Gauss-Bonnet 
gravity \cite{Padmanabhan02, Padmanabhan05, 
ParanjapeSarkarPadmanabhan06, 
KothawalaSarkarPadmanabhan07, NozawaMaeda08}.

The first law of black hole thermodynamics for event horizons in 
general relativity is 
\begin{equation}
T\delta S= \delta M -\Omega_H \delta J + \, ... \,,
\end{equation} 
where $T$ and $S$ are the horizon temperature and entropy, $M$ 
and $J$ are the hole mass and angular momentum measured at 
spatial infinity, and $\Omega_H$ is the angular velocity of the 
horizon, while the ellipsis denote extra terms which appear if 
the black hole possesses other charges. The first law relates the 
quantities $M$ and $J$ 
measured at infinity  with the local quantities $ S, T, A$, and $ 
\Omega_H$ on the horizon. This property of the first law still 
holds true  
in alternative theories of gravity, as emphasized in 
\cite{JacobsonKangMyers94}, but the expression of the entropy 
$S_{BH}$ must be changed in these theories. The fact that the 
expression $S_{BH}=\frac{A}{4G}$ does not hold in alternative 
gravity has been known since the 1980s 
\cite{RaychaudhuriBagchi83,  CallanMyersPerry88, MyersSimon88, 
LuWise93, Visser93b, CreightonMann95, Myersproceedings}.

Various techniques have been developed to compute black hole 
entropy, including Wald's Noether charge method 
\cite{Wald93, IyerWald94, IyerWald95, Visser93b, 
JacobsonKangMyers94},  field 
redefinition techniques \cite{JacobsonKangMyers94}, and the 
Euclidean path integral approach \cite{GibbonsHawking77}. Wald's 
Noether charge method  relies on a Lagrangian formulation 
of the first law of  black hole thermodynamics and is applicable 
to stationary black  holes  with bifurcate Killing horizons in 
any relativistic theory of gravity  with  diffeomorphism 
invariance in 
arbitrary spacetime dimension. This method  was applied to black 
holes in Palatini modified gravity \cite{Vollick07}, metric  
$f(R)$ 
gravity,  and in  other gravitational theories 
\cite{BrisceseElizalde08, Brusteinetal09}.

Due to the recent interest in modified gravity coming from 
cosmology, in the following we focus on the  Bekenstein-Hawking 
entropy \cite{Bekenstein73, Hawking75} in 
scalar-tensor and modified gravity. Before proceeding, we recall 
the equivalence between metric and Palatini $f(R)$ gravity and 
scalar-tensor theories in the next subsection. This discussion 
also serves  the purpose of establishing the notations used 
in the following sections.

\subsection{Metric and Palatini $f(R)$ gravity as scalar-tensor 
theories}

The equivalence between  metric and Palatini 
$f(R)$ theories and scalar-tensor gravity has been discussed 
and rediscovered many times  \cite{STequivalence1, 
STequivalence2, STequivalence3, STequivalence4, STequivalence5, 
STequivalence6}. 
The action of metric $f(R)$ gravity is \cite{CCT1, CCT2, CDTT} 
\begin{equation}  \label{actionmetric}
I_{metric}=\frac{1}{2\kappa} \int d^4 x \, \sqrt{-g} \, 
f(R)+I^{(matter)} \,,
\end{equation}
where $\kappa=8\pi G$. Variation with respect to the (inverse) 
metric tensor  $g^{\mu\nu}$ yields 
the field equations
\begin{equation} \label{metricfieldeqs}
f'(R)R_{\mu\nu}-\frac{f(R)}{2} \, 
g_{\mu\nu}=\nabla_{\mu}\nabla_{\nu} f'(R) -g_{\mu\nu} \Box 
f'(R) +\kappa\, T_{\mu\nu} \,,
\end{equation}
where a prime denotes differentiation with respect to $R$.  
If  $f''(R) \neq 0$, introduce the  auxiliary scalar field $\phi 
=R$ and consider the action
\begin{equation} \label{equivalentmetric}
I = \frac{1}{2\kappa}  \int d^4 x \, \sqrt{-g} \left[ \psi( 
\phi)R -V(\phi)  \right] + I^{(matter)} \,,
\end{equation}
where 
\begin{equation}
\psi(\phi) = f'(\phi) \,, \;\;\;\;\;\;
V(\phi)=\phi f'(\phi)-f(\phi) \,.
\end{equation}
This action  reduces trivially  
to~(\ref{actionmetric})  if $\phi=R$. 
{\em Vice-versa}, varying~(\ref{equivalentmetric}) with  respect 
to 
$g^{\mu\nu}$ gives
\begin{equation}
G_{\mu\nu}=\frac{1}{\psi}\left( \nabla_{\mu}\nabla_{\nu} \psi-  
g_{\mu\nu}\Box\psi 
-\frac{V}{2}\, g_{\mu\nu} \right)+\frac{\kappa}{\psi} \, 
T_{\mu\nu}  \,,
\end{equation}
while the variation with respect to $\phi$ yields
\begin{equation}
R\, \frac{d\psi}{d\phi} -\frac{dV}{d\phi}=\left( R-\phi 
\right)f''(\phi)=0
\end{equation}
and $\phi=R$ under the assumption  
$f''\neq 0$ (in fact, local stability 
requires $f''(R)>0$ \cite{DolgovKawasaki03, mattmodgrav1, 
mattmodgrav2,  Odintsovconfirm, review}). Therefore,  
the (massive) scalar  field  $\phi=R$  is a dynamical degree of 
freedom which   
satisfies the trace equation
\begin{equation}
3f''(\phi)\Box \phi+3f'''(\phi)\nabla^{\sigma} 
\phi\nabla_{\sigma} \phi +\phi 
f'(\phi)  -2f(\phi) =\kappa \, {T^{\mu} }_{\mu}\,.
\end{equation}
It is more convenient to consider  $ \psi \equiv f'(\phi)$ 
instead of  $\phi$; then, $\psi$ satisfies the equation 
\begin{equation}
3\Box \psi +2 U(\psi) -\psi\, \frac{dU}{d\psi}=\kappa \, 
{ T^{\mu} }_{\mu} 
\end{equation}
with $ U(\psi)=V(\phi(\psi))-f(\phi(\psi)) $, and the action 
\begin{equation}
I =\frac{1}{2\kappa} \int d^4 x \, \sqrt{-g} \left[ \psi R 
-U(\psi) \right] + I^{(matter)} \,,
\end{equation}
is clearly that of a Brans-Dicke theory with Brans-Dicke 
parameter $\omega=0$ \cite{BransDicke}.

In the Palatini approach, both the metric $g_{\mu\nu}$ and the 
connection  $\Gamma^{\alpha}_{\mu\nu}$ are treated as independent 
variables, {\em i.e.}, the  connection is not the metric 
connection of $g_{\mu\nu}$. The metric and Palatini variations 
produce the same field equations in GR \cite{WaldGRbook} and in 
Lovelock 
gravity \cite{Exirifard} but not for general Lagrangians 
non-linear in $R$.

The Palatini action 
\begin{equation}\label{actionPalatini}
I _{Palatini} =\frac{1}{2\kappa}\int d^4 x \, \sqrt{-g} \, 
f\left(  \cal{R}  \right) + I^{(matter)}\left[ g_{\mu\nu}, 
\psi^{(m)} \right] 
\end{equation}
implicitly contains two Ricci  tensors: the usual $R_{\mu\nu}$  
constructed with the metric connection of  the   physical metric 
$g_{\mu\nu}$, and $\cal{R}_{\mu\nu}$ constructed with the  
non-metric connection $\Gamma^{\alpha}_{\mu\nu}$. 
$\cal{R}_{\mu\nu} $ gives rise to the scalar ${ \cal R}\equiv 
g^{\mu\nu}\cal{R}_{\mu\nu}$.  The matter Lagrangian  does 
not depend explicitly from the connection $\Gamma$, but only  
from the metric and the matter fields $\psi^{(m)}$.

The variation of the Palatini action~(\ref{actionPalatini}) 
produces the field equation
\begin{equation} \label{Palatinifieldeq1}
f'({\cal R}) {\cal R}_{\mu\nu}-\frac{ f( {\cal R})}{ 2} \, 
g_{\mu\nu}=\kappa \, 
T_{\mu\nu} \,,
\end{equation}
in which there are no  second covariant derivatives of $f'$,  in 
contrast  with eq.~(\ref{metricfieldeqs}). Varying with respect 
to the independent connection yields
\begin{equation} \label{Palatinifieldeq2}
\bar{\nabla}_{\alpha} \left( \sqrt{-g} \, f'( {\cal R}) 
g^{\mu\nu} \right)
-\bar{\nabla}_{\sigma} \left( \sqrt{-g} \, f'({ \cal R}) 
g^{\sigma ( \mu }\right) \delta^{\nu )}_{\alpha} =0 \,,
\end{equation}
where $\bar{\nabla}_{\mu}$ denotes the covariant derivative of   
the non-metric connection $\Gamma$. The trace of 
eqs.~(\ref{Palatinifieldeq1}) and (\ref{Palatinifieldeq2}) 
yields
\begin{equation} \label{Palatinitrace}
f'({\cal R}) {\cal R} -2f( {\cal R})=\kappa \, {T^{\mu}}_{\mu}  
\end{equation}
and
\begin{equation} \label{Palatinifieldeq3}
\bar{\nabla}_{\gamma} \left( \sqrt{-g}  \, f'( {\cal R}) 
g^{\mu\nu} \right)=0 \,.
\end{equation}
$f'( {\cal R})$  is non-dynamical, contrary to the scalar 
degree of freedom of metric $f(R)$ 
gravity.   It is possible to eliminate completely the non-metric 
connection from  the field equations, which are then rewritten as
\begin{eqnarray}
 G_{\mu\nu} & = & \frac{\kappa}{f'}\, 
T_{\mu\nu}-\frac{1}{2}\left( 
R-\frac{f}{f'}\right) g_{\mu\nu} +\frac{1}{f'} 
\left(\nabla_{\mu}\nabla_{\nu} 
-g_{\mu\nu}\Box \right) f'\nonumber\\
&&\nonumber \\
& - & \frac{3}{2(f')^2}  \left[  
\nabla_{\mu} f' \nabla_{\nu} 
f' -\frac{1}{2} \, g_{\mu\nu} \nabla_{\gamma} f' \nabla^{\gamma} 
f' \right] \,.
\label{Palatinireformulated}
\end{eqnarray}

To see the equivalence with a  Brans-Dicke theory, proceed as  
in the metric formalism:  introduce    $\phi  ={ \cal R}$ and 
$\psi \equiv f'(\phi)$ in the  action~(\ref{actionPalatini}). 
Apart from a  boundary term which  can be discarded, the action 
is rewritten in terms of $g_{\mu\nu}$ and $R_{\mu\nu}$ as (see 
\cite{review} and the references therein for details)
\begin{equation} \label{equivalentPalatini}
I_{Palatini}=\frac{1}{2\kappa}\int d^4x \, \sqrt{-g} \left[ \psi 
R  +\frac{3}{2\psi} \, \nabla^{\mu} \psi\nabla_{\mu} \psi 
-V(\psi) \right] + I^{(matter)} \,.
\end{equation}
This is the  action of  a Brans-Dicke  theory with Brans-Dicke 
parameter $\omega=-3/2$ and non-vanishing potential for $\psi$ 
\cite{BransDicke}.

%%%%%%%%%%%%%%%%%%%%%%%%%%%%%%%%%%%%%%%%%%%%%%%%%%%%%%%%%%%%
\section{Scalar-tensor gravity}

Black hole entropy in Brans-Dicke gravity was analyzed by Kang 
\cite{Kang96} following numerical  studies 
demonstrating that
during the collapse of dust to black holes in this 
class  of theories the area law  valid in Einstein's theory 
({\em i.e.}, the horizon area can never decrease) is violated 
\cite{ScheelShapiroTeukolsky1, ScheelShapiroTeukolsky2, 
Kerimo1, Kerimo2}. 
Kang realized that the problem is not in the area law itself but 
rather in the expression of the black hole entropy, which is not 
simply one quarter of the area in these theories. The expression 
for the entropy is rather
\begin{equation} \label{Kangentropy}
S_{BH}=\frac{1}{4} \int_{\Sigma} d^2x \sqrt{g^{(2)} } \, \phi  
=\frac{\phi A}{4}
\,,
\end{equation}
where $\phi$ is the Brans-Dicke scalar and $g^{(2)}$ is the 
determinant of the restriction $g_{\mu\nu}^{(2)} \equiv 
g_{\mu\nu} \left.\right|_{\Sigma}$ of the metric 
$g_{\mu\nu}$ to the horizon surface $\Sigma$. This expression can 
be 
understood by the simple replacement of the Newton constant 
$G$ with the effective gravitational coupling 
\begin{equation}
G_{eff}=\phi^{-1}
\end{equation}
in Brans-Dicke theory \cite{Kang96} so that $S_{BH}=A/4G_{eff}$. 
The quantity $S_{BH}$ is 
non-decreasing. This philosophy of replacing the gravitational 
coupling with the effective gravitational coupling that would 
appear if one were to rewrite the field equations of the theory 
as effective Einstein equations and read scalar field (or, in 
$f(R)$ gravity, geometric)  terms as an effective form of 
matter, carries over to  more general scalar-tensor gravities 
and to other gravitational theories. The 
expression~(\ref{Kangentropy}) has now been derived 
using various procedures \cite{JacobsonKangMyers94, IyerWald94, 
Visser93b}.

Following \cite{Kang96}, one can also consider the Einstein  
frame representation of  Brans-Dicke theory given by the conformal 
rescaling of the metric
\begin{equation}
g_{\mu\nu}\longrightarrow \tilde{g}_{\mu\nu} \equiv \Omega^2 \,  
g_{\mu\nu} \,, \;\;\;\; \Omega=\sqrt{ G\phi} \,,
\end{equation}
accompanied by the scalar field redefinition $\phi \rightarrow 
\tilde{\phi}$ with $\tilde{\phi}$ given by
\begin{equation}
d\tilde{\phi}= \sqrt{ \frac{2\omega +3}{16\pi G}} \,  
\frac{d\phi}{\phi} \,.
\end{equation}
The Brans-Dicke action \cite{BransDicke}
\begin{equation}
I_{BD}=\int d^4x \, \frac{ \sqrt{-g}}{16\pi}  \left[ \phi R 
-\frac{\omega}{2} \,  g^{\mu\nu} \nabla_{\mu}\phi 
\nabla_{\nu}\phi -V(\phi)  +{\cal  L}^{(m)} \right]
\end{equation}
assumes the Einstein frame form
\begin{equation}
I_{BD}=\int d^4x \,  \sqrt{-\tilde{g}}   \left[  
\frac{ \tilde{R} }{ 16\pi G}  
-\frac{1}{2} \,  \tilde{g}^{\mu\nu} \tilde{\nabla}_{\mu}
\tilde{\phi}  \tilde{\nabla}_{\nu} \tilde{\phi} -U( \tilde{\phi})  
+ \frac{ {\cal  L}^{(m)} }{\left( G\phi \right)^2}  \right] \,,
\end{equation}
where a tilde denotes Einstein frame (rescaled)  quantitites and 
\begin{equation}
U \left( \tilde{\phi} \right)=\frac{ V(\phi( \tilde{\phi}) )}{ 
\left( G\phi ( \tilde{\phi}) \right)^2} 
\end{equation}
with $\phi=\phi( \tilde{\phi})$.
In the Einstein frame the gravitational coupling is a true  
constant but matter couples explicitly to the scalar field and 
what were massive test particles following timelike geodesics of 
the 
Jordan frame metric $g_{\mu\nu}$ do not  follow 
geodesics of the rescaled metric $\tilde{g}_{\mu\nu}$. Null 
geodesics are left unchanged by the conformal rescaling, as well 
as null vectors and all forms of conformally invariant matter. 
A black hole event horizon, being a null surface, is also 
unchanged. The area of an event horizon is not, and the change 
in the entropy formula 
$ S_{BH}=\frac{A}{4G} \rightarrow  \frac{A}{4G_{eff}}=\frac{\phi 
A}{4}  $ can be understood as the  change in the area due to the 
conformal rescaling of $g_{\mu\nu}$. In fact, $ 
\tilde{g}_{\mu\nu}^{(2)} = \Omega^2 \, 
g_{\mu\nu}^{(2)} $ and, since the horizon surface is unchanged, 
the Einstein frame area is
\begin{equation}
\tilde{A}=\int_{\Sigma}d^2x \, \sqrt{ \tilde{g}^{(2)}}=
\int_{\Sigma} d^2x \,\Omega^2 \,  \sqrt{ g^{(2)}}=
G \phi \,A 
\end{equation}
assuming that the scalar field is constant on the horizon (if 
this is not true the surface gravity is unlikely to be constant 
according to any sensible definition and the zeroth law of black 
hole thermodynamics fails). Therefore, the entropy-area relation  
$ \tilde{S}_{BH}=\tilde{A}/4G $  still holds in the 
Einstein frame. 
This should be expected since, {\em in vacuo}, the theory reduces 
to 
general relativity with varying units of length $ \tilde{l}_u 
\sim \Omega \, l_u$, time $\tilde{t}_u \sim \Omega \, t_u $, and 
mass $\tilde{m}_u=\Omega^{-1} \, m_u $  (where $t_u, l_u$, and 
$m_u$ are the constant units of time, length, and mass in the 
Jordan frame, respectively), and derived 
units vary 
as well \cite{Dicke}. Then, an area scales as $A \sim \Omega^2 = 
G\phi $ and, in units in which $c=\hbar= 1$ the entropy is 
dimensionless and is not rescaled. Therefore, the Jordan frame 
and Einstein frame entropies coincide (a point noted in 
\cite{Kang96}):
\begin{equation}
\tilde{S}=\frac{ \tilde{A}}{4G_{eff}}=\frac{A}{4G}=S \,.
\end{equation}
The equality between black hole entropies in the Jordan and 
Einstein frames is not restricted to scalar-tensor gravity but 
extends   to all theories with action $\int d^4x \sqrt{-g} \, 
f\left(  g_{\mu\nu}, R_{\mu\nu}, \phi, \nabla_{\alpha}\phi 
\right) $ which admit an Einstein frame representation 
\cite{KogaMaeda98}.

A consequence of this equivalence which is worth noting is that 
the Jordan and the Einstein frames turn out to be  physically 
equivalent again. A debate on whether these two frames 
are physically equivalent has been going on for years and the  
issue still causes frequent confusion. It seems now established 
(although many authors may disagree with this statement) that, 
at the classical level, the two frames are merely different 
representations of the same physics (see the discussion in
\cite{Dicke, Flanagan, FaraoniNadeauconfo}). There are 
potential problems 
arising from the fact that many fundamental properties of 
physical theories, including the cherished Equivalence 
Principle, turn out to be dependent on the conformal 
representation adopted, which means that the fundamental 
properties of gravitational theories should be reformulated in a 
representation-independent manner  (\cite{ThomasStefanoValerio} 
and references therein).  The classical  equivalence is expected 
to break down at the quantum level, in 
the same way that the quantization of Hamiltonians related by 
canonical transformations produce inequivalent energy spectra 
and eigenfunctions \cite{inequivalentHamiltonians1, 
inequivalentHamiltonians2, inequivalentHamiltonians3}. While 
quantum gravity is expected to definitely break the equivalence 
between conformal frames, the situation is not so clear at the 
semiclassical level \cite{Flanagan}. Black hole thermodynamics 
is not purely classical: what ultimately makes it meaningful  
is the discovery  of Hawking radiation, a semiclassical 
phenomenon. It is widely believed that 
black holes open a 
window onto quantum gravity and therefore, implicitely, that 
at least some aspects of their thermodynamics will be preserved 
and derived theoretically in the quantum gravity regime. It is 
therefore significant that 
the physical equivalence between conformal frames holds for 
(semiclassical)  black hole entropy. Whatever gravity knows about 
black hole thermodynamics and the underlying microscopic 
statistical mechanics, it seems to know also about the 
equivalence between  conformal frames.

Another consequence of the discussion above using 
the Einstein frame representation of Brans-Dicke (and, 
by extension, of more general scalar-tensor) gravity is that if  
the  scalar field vanishes on  the horizon of a Brans-Dicke black 
hole  the latter is attributed zero temperature and, in the 
light of the previous considerations, zero entropy. These black 
holes with vanishing $\phi$, dubbed ``cold black holes'',  
have been the subject of a not insignificant amount of literature  
\cite{CampanelliLousto1, CampanelliLousto2, coldblackholes1, 
coldblackholes2, coldblackholes3, coldblackholes4, 
coldblackholes5, coldblackholes6, coldblackholes7, 
coldblackholes8, coldblackholes9}.

If $\phi$ diverges on the horizon of scalar-tensor black holes, 
the entropy is infinite there, which  seems to 
rule out the possibility that $\phi \rightarrow \infty$ as the 
horizon is approached. Interestingly, scalar hair always seems 
to vanish or diverge on the black hole horizon, making  
the no-hair theorems all the more plausible in scalar-tensor 
gravity. On the other hand, in the context of Brans-Dicke theory 
(with zero scalar field potential), there is a well-known theorem 
by Hawking \cite{HawkingBD} stating that, barring the 
situations of scalar field vanishing or diverging on the horizon, 
all stationary Brans-Dicke black holes are the same as in general 
relativity in the sense that the scalar field becomes constant 
outside the horizon and, in this situation, the theory reduces 
to general relativity. (Note that the limit of scalar-tensor 
gravity to general relativity as $\phi$ becomes constant and the 
explicit dependence of $\phi$ on  the  Brans-Dicke parameter 
$\omega$ in this limit are not entirely trivial 
\cite{GRlimit1, GRlimit2, GRlimit3, 
GRlimit4, GRlimit5, GRlimit6, GRlimit7, 
GRlimit8, GRlimit9, GRlimit10, 
GRlimit11, ScheelShapiroTeukolsky2}.) This conclusion is 
corroborated by  numerical studies of black hole collapse in 
Brans-Dicke gravity  \cite{ScheelShapiroTeukolsky1, 
ScheelShapiroTeukolsky2, 
Kerimo1, Kerimo2}. 
According to the  
discussion above, it seems that these black holes should be  
discarded as 
pathological from the thermodynamical point of view, which makes 
the black holes of general relativity the only possible final 
state of equilibrium in Brans-Dicke theory. (The use of entropic 
considerations to select correct gravity theories among the 
class of metric $f(R)$ models is advocated, {\em e.g.},  in 
\cite{Kim97, BrisceseElizalde08}.)   
It seems that, discarding cold black holes and those with 
diverging horizon entropy, Hawking's  theorem 
\cite{HawkingBD} should be extendible  to all scalar-tensor 
theories (work is in progress on this subject).

%%%%%%%%%%%%%%%%%%%%%%%%%%%%%%%%%%%%%%%%%%%%%%%%%%%%%%%%%%%%
\section{Metric $f(R)$ gravity}

It is by now well known that the area formula $S_{BH}=A/4G$  
gets corrected as  
\begin{equation}\label{metricf(R)entropy}
S_{BH}=\frac{ f'(R)A}{4G}
\end{equation}
in metric $f(R)$ gravity \cite{BrevikNojiriOdintsovVanzo04, 
CognolaElizaldeNojiriOdintsovZerbini05, AkbarCaiPLB06, 
GongWangPRL07,  Brusteinetal09}. 

The Noether charge method was applied  to metric $f(R)$ gravity 
on various occasions \cite{BrevikNojiriOdintsovVanzo04, 
CognolaElizaldeNojiriOdintsovZerbini05, Brusteinetal09}, usually  by 
assuming that the black hole configuration is static and working 
in $D$ spacetime dimensions. A common result of these studies is  
that the  usual entropy-area relation  is still valid provided 
that 
Newton's constant $G$ is replaced by a suitable effective 
gravitational coupling $G_{eff}$. The identification of $G_{eff}$  
with $G/f'(R)$  in metric $f(R)$ gravity is 
straightforward based on inspection of the action, or of the 
field equations of the theory rewritten in the form of effective 
Einstein equations. Those who find this identification too naive 
should look at the recent work 
of Brustein and collaborators \cite{Brusteinetal09}, in which 
the   identification of  $G_{eff}$ is made by using the 
matrix 
of  coefficients of the kinetic terms 
for metric perturbations \cite{Brusteinetal09}. The metric 
perturbations contributing to the Noether charge in Wald's 
formula and its generalizations are identified with specific 
metric perturbation polarizations associated with fluctuations of 
the  area density on the bifurcation surface $\Sigma$ of the 
horizon (this is the $(D-2)$-dimensional spacelike cross-section 
of a Killing horizon on which the Killing field vanishes, and 
coincides with the intersection of the two null hypersurfaces 
comprising this horizon). The horizon entropy is 
\begin{equation} 
S_{BH}=\frac{A}{4G_{eff}} 
\end{equation}
for a theory described by the action
\begin{equation} \label{Brusteinaction}
I =\int d^4x \, \sqrt{-g} \, {\cal L} \left( g_{\mu\nu}, 
R_{\alpha\beta\rho\lambda}, 
\nabla_{\sigma} R_{\alpha\beta\rho\lambda}, \phi, 
\nabla_{\alpha}\phi, \, ...\right) \,,
\end{equation}
where $\phi$ is  a gravitational scalar field.  The Noether 
charge is
\begin{equation}
S=-2\pi \int_{\Sigma} d^2x\, \sqrt{g^{(2)} }  \left( \frac{ 
\delta {\cal 
L}}{\delta 
R_{\mu\nu \rho\sigma}} \right)_{(0)} { \hat{ {\mathbf \epsilon}}} 
_{\mu\nu}  
{\hat{ {\mathbf \epsilon} }}_{\rho\sigma}  \,,
\end{equation}
where ${\hat{ {\mathbf \epsilon}}}_{\rho\sigma}   $ is the 
(antisymmetric) binormal vector  
to the bifurcation surface $\Sigma$ normalized to $
{\hat{ {\mathbf  \epsilon}}}^{ab}  
{\hat{ {\mathbf \epsilon}}}_{ab}  =-2$ and the subscript $(0)$ 
denotes the fact that the quantity in brackets is evaluated on 
solutions of 
the equations of motion (on the bifurcation 
surface $\Sigma$, the binormal satisfies $\nabla_{\mu} 
\chi_{\nu}={ \hat{  {\mathbf \epsilon} }}_{\mu\nu}$, where 
$\chi^{\mu}$ is the Killing 
field vanishing on the horizon). The effective gravitational 
coupling is 
then calculated to be \cite{Brusteinetal09}
\begin{equation}
G_{eff}^{-1} =-2\pi \left( \frac{ \delta {\cal L}}{\delta 
R_{\mu\nu\rho\sigma}} \right)_{(0)} {\hat{ {\mathbf  
\epsilon}}}_{\mu\nu}  
 {\hat{ {\mathbf \epsilon}} }_{\rho\sigma} \,.
\end{equation}
These prescriptions apply not only to $f(R)$ gravity but to 
other theories described by the action~(\ref{Brusteinaction}) as 
well.

For metric $f(R)$ gravity described by the Lagrangian density 
${\cal L}=f(R)$ this prescription yields $
G_{eff}=G/f'(R)$ and the entropy~(\ref{metricf(R)entropy}).  
This calculation is consistent with the description of metric 
$f(R)$ gravity as  a scalar-tensor theory with a massive 
scalar degree of freedom $f'(R)$ and with the 
corresponding eq.~(\ref{Kangentropy}) of scalar-tensor gravity.

%%%%%%%%%%%%%%%%%%%%%%%%%%%%%%%%%%%%%%%%%%%%%%%%%%%%%%%%%%%%
\section{Palatini $f(R)$ gravity}

Wald's Noether charge method can be applied also to 
Palatini $f( {\cal R})$ gravity. In a slightly different  
notation from  the one used earlier,  the  entropy  of a black 
hole static  horizon is given by the (local) Noether charge
\begin{equation}\label{PalatiniNoethercharge}
S_{BH}=\frac{2\pi}{\kappa_g} \int_{\Sigma} {\mathbf Q} \,,
\end{equation}
where the $(D-2)$-form $ {\mathbf  Q} $ is the Noether potential 
associated with 
the 
diffeomorphisms of the spacetime manifold, $\Sigma$ is the 
bifurcation surface of the black hole, and $\kappa_g$ is the 
surface gravity  on the horizon (the 
entropy~(\ref{PalatiniNoethercharge}) 
does not change when calculated on any cross-section of the 
horizon \cite{JacobsonKangMyers94}).   Vollick \cite{Vollick07} 
considered $D$-dimensional Palatini $f(R)$ gravity described by 
the action 
\begin{equation}
I_{Palatini}=\int d^Dx \sqrt{-g} \left[ \frac{f({\cal R})}{16\pi 
G} +{\cal L}^{(m)} \right] \,.
\end{equation}
Palatini $f( {\cal R})$ gravity {\em in vacuo} is equivalent to 
general 
relativity with a  cosmological constant and the entropy of a  
stationary black hole is found to be 
\begin{equation}
S_{BH}= \frac{ f'({\cal R}) A}{4G} \,.
\end{equation}
In the presence of matter it is useful to consider the 
trace~(\ref{Palatinitrace}) of the field equations which is an 
algebraic (or trascendental) equation, not a 
differential equation. This fact reflects the non-dynamical 
nature of the scalar  $f'({\cal R})$ present in  the theory 
and has been emphasized many times in the literature (cf. the 
references in \cite{review, DeFeliceTsujikawa10}).  Using 
eq.~(\ref{Palatinitrace}), 
when the trace $ {T^{\mu}}_{\mu}$ is constant (and, in 
particular, for conformally invariant matter for which 
$ {T^{\mu}}_{\mu} =0 $), it is possible to eliminate the 
Ricci curvature ${\cal R}$ in terms of $ {T^{\mu}}_{\mu}$, which 
becomes a constant and also $f'( {\cal R} )$ is 
constant. 
Then, the theory is again equivalent to general relativity with a 
cosmological constant, as described by the field equations which 
can be rewritten as \cite{Vollick07}
\begin{equation}
G_{\mu\nu}[ g_{\alpha\beta} ] =\frac{8\pi G}{f'}\, T_{\mu\nu} 
-\left( \frac{D-2}{2D} \right) \, R g_{\mu\nu} \,.
\end{equation}
Recasting the field equations in this way allows one to identify 
the effective gravitational coupling of the theory
\begin{equation}
G_{eff}=\frac{G}{f'( {\cal R})}
\end{equation}
and consequently the black hole entropy given by the Noether 
charge corresponds simply to the familiar expression with 
Newton's constant $G$  replaced by $G_{eff}$ or \cite{Vollick07}
\begin{equation}
S_{BH}=\frac{A}{4G_{eff}}=\frac{f'( {\cal R}) A}{4G} \,.
\end{equation}

In the presence of matter with non-constant trace 
${T^{\mu}}_{\mu}$ the situation is more complicated  and the 
black hole entropy depends on the ratio of the effective 
gravitational couplings on the horizon and at spatial infinity:
\begin{equation}
S_{BH}= \frac{ f'_{\Sigma} }{ f'_{\infty}} \, \frac{A}{4G} \,,
\end{equation}
where $f'_{\Sigma}$ is the value of $f'( {\cal R})$ on the 
horizon and $ f'_{\infty}$ is the value far away from the black 
hole \cite{Vollick07}.

%%%%%%%%%%%%%%%%%%%%%%%%%%%%%%%%%%%%%%%%%%%%%%%%%%%%%%%%%%%%
\section{Dilaton gravity (metric and Palatini)}

Theories with action 
\begin{equation}
I =\int d^D x \, \frac{ \sqrt{-g} }{16\pi G} \,  f \left( 
g^{\mu\nu}, R_{(\mu\nu)} \right) 
\end{equation}
were considered in Ref.~\cite{Vollick07}. The variation with 
respect 
to $g^{\mu\nu}$ yields the field equations
\begin{equation} 
\frac{ \partial f}{\partial g^{\mu\nu} } -\frac{f}{2}\, 
g_{\mu\nu}=0 \,,
\end{equation}
while the variation with respect to the independent connection 
$\Gamma$ yields
\begin{equation}
\bar{\nabla}_{\alpha} \left[ \sqrt{-g} \,\, \frac{\partial f}{ 
\partial {\cal 
R}_{(\mu\nu)} } \right]=0 \,.
\end{equation}
Again, the {\em vacuum}  theory  is equivalent to general 
relativity 
with an effective  cosmological constant, {\em i.e.}, 
the effective matter  tensor due to the geometric terms has the 
form 
\begin{equation}
\frac{\partial f}{ \partial {\cal  R}_{(\mu\nu)}}=\lambda 
g_{\mu\nu} \,,
\end{equation}
where $\lambda$ is a constant, the effective gravitational 
coupling is $G_{eff}=G/\lambda$, and  the black hole entropy is 
\cite{Vollick07}
\begin{equation}
S_{BH}=\frac{A}{4G_{eff}} =\frac{\lambda A}{4G} \,.
\end{equation}

A dilaton gravity in the Palatini approach, described by
\begin{equation}
I =\int d^D x \sqrt{-g} \, \mbox{e}^{-2\phi} \left[ 
\frac{f( {\cal R} )}{16\pi G} -\frac{\alpha}{2}\, g^{\mu\nu} 
\nabla_{\mu}\phi \nabla_{\nu}\phi \right] \,,
\end{equation}
was also studied in Ref.~\cite{Vollick07}. Here  $\alpha$ is a 
constant (corresponding to $\sim G^{-1}$ in 
string theory) and $\phi$ is the dilaton field. The equations of 
motion 
\begin{eqnarray}
&& f'\left( {\cal R} \right)  {\cal R}_{(\mu\nu)} -\frac{f 
\left( {\cal R}\right) }{2}\, g_{\mu\nu} =8\pi G 
\alpha 
\left( \nabla_{\mu}\phi\nabla_{\nu}\phi -\frac{1}{2}\, g_{\mu\nu}
\nabla^{\alpha}\phi\nabla_{\alpha}\phi \right) \,,\\
&&\nonumber\\
&& \bar{\nabla}_{\alpha}\left( \sqrt{-g} \, \mbox{e}^{-2\phi} \, 
f' 
g^{\mu\nu} \right)=0 \,,
\end{eqnarray}
lead to the identification of the effective gravitational 
coupling $G_{eff}=G/f'_{\infty}$ and the Noether charge entropy 
is \cite{Vollick07}
\begin{equation}
S_{BH}= \left[ \mbox{e}^{-2\phi} f' \right]_{\Sigma} \, 
\frac{A}{4G}=\left. 
\mbox{e}^{-2\phi} \right|_{\Sigma} \, \frac{ f'_{\Sigma}}{
f'_{\infty}} \, \frac{A}{4G_{eff}}  \,.
\end{equation}

A similar dilaton gravity in the metric formalism, with action
\begin{equation}
I =\int d^Dx \sqrt{-g} \, \, \frac{ \mbox{e}^{-2\phi} R}{16 \pi 
G} 
\end{equation}
was considered in Ref.~\cite{Brusteinetal09}. In spite of the 
differences  between the Palatini and the metric formalisms, the 
result is the same: Brustein and collaborators find again 
\cite{Brusteinetal09}  
\begin{equation}
S_{BH} =   
\frac{\mbox{e}^{-2\phi} \left. \right|_{\Sigma} A}{4G}  \,.
\end{equation}

%%%%%%%%%%%%%%%%%%%%%%%%%%%%%%%%%%%%%%%%%%%%%%%%%%%%%%%%%%%%
\section{Conclusions}

The entropy-area relation $S=A/4G$ familiar from general 
relativity is still valid in both metric and Palatini $f(R)$ 
gravities and in the much larger class of scalar-tensor theories 
provided that Newton's constant is replaced by a suitable 
effective gravitational coupling strength $G_{eff}$ 
({\em i.e.}, $\phi^{-1}$ in 
scalar-tensor gravity, or $G/f'$ in modified gravities). Black 
hole entropy has been studied in quantum gravity-inspired 
theories  that depart more radically than scalar-tensor ones from  
Einstein theory: for  example, entropy in Horava-Lifshitz 
gravity is given by a more complicated expression  
\cite{MyungKim09}.  The  thermodynamics of cosmological horizons 
is also of  interest, see 
Refs.~\cite{CognolaElizaldeNojiriOdintsovZerbini05,  Wang05,
AkbarCaiPLB06, GongWangPRL07, MosheniSadjadi07, 
CaiCaoNPB07, CaiCaoPRD07, CaiCaoHuJHEP08} for discussions in the 
context of metric  $f(R)$ gravity.  We 
have not discussed studies of the Bekenstein-Hawking entropy in 
Lovelock  \cite{MyersSimon88, JacobsonMyers93, 
ParanjapeSarkarPadmanabhan06, CaiPLB04, 
Correa05} and Gauss-Bonnet \cite{Padmanabhan02, Padmanabhan05, 
ParanjapeSarkarPadmanabhan06, 
KothawalaSarkarPadmanabhan07} gravity, 
or in  theories with Lorentz 
violation in which thermodynamical considerations have been 
claimed to allow for the 
possibility of perpetual motion machines of the second kind 
\cite{DubovskySibiryakov06}---see also 
\cite{ElingFosterJacobsonWall07, JacobsonWall08}. This claim  
has been reconsidered and shown to be invalid in 
tensor-vector-scalar (TeVeS) theories in 
Ref.~\cite{SagiBekenstein08}. This debate shows once again that 
black 
hole thermodynamics in alternative  theories of gravity can be 
quite interesting and may be used in the future to constrain the 
class of effective actions inspired by low-energy quantum 
gravity.

\section*{Acknowledgments}

The author thanks Remo Garattini for his invitation to 
contribute to this volume and the Natural Sciences and  
Engineering Research Council of Canada (NSERC) for financial 
support.

\end{document}